\documentclass[12pt]{article}
																																																																													\pdfoutput=1																																																																											 
\usepackage{amsmath,amssymb}
\usepackage{graphicx}
\usepackage[pdftex]{hyperref}
\DeclareGraphicsExtensions{.eps,.bmp,.wmf,.jpeg,.pdf}
\numberwithin{equation}{section}
\def\be{\begin{equation}}
\def\ee{\end{equation}}

\def\bea{\begin{eqnarray}}
\def\eea{\end{eqnarray}}
\title{Finite scale factor and future singularities}
\author{L.N. Granda\thanks{luis.granda@correounivalle.edu.co} \\ {\small\it Departamento de Fisica, Universidad del Valle}\\{\small\it A.A. 25360, Cali, Colombia}}

\date{}
\begin{document}
\maketitle

\begin{abstract}
\noindent The main characteristic of the dark energy is its negative pressure. In a homogeneous and isotropic FRW background, we consider several models for the dark energy fluid, which lead to finite time future singularities of the type I-IV, by introducing the pressure density as a function of the scale factor. This approach gives acceptable behavior of the dark energy equation of state. We give various numerical examples of models with type I-IV singularities, that show very similar late time behavior, making it difficult to determine the type of singularity that would take place in the future.\\

\end{abstract}

\maketitle

\section{Introduction}
\noindent 
According to current astrophysical data \cite{riess}, \cite{perlmutter}, \cite{kowalski}, \cite{hicken}, \cite{komatsu}, \cite{percival}, \cite{komatsu1}, there is the  possibility that the dark energy (DE) component of the universe is causing an accelerated expansion that goes beyond the acceleration given by the cosmological constant. These observations allow to conclude that the DE equation of state (EoS) $w$ lies in a narrow interval around $w=-1$, being quite consistent with values bellow this limit. This opens the possibility to an universe that undergoes three phases; namely, the initial matter dominance phase with decelerated expansion, with transition to accelerated phase $-1<w<-1/3$, and the probably super-accelerated expansion, called phantom phase characterized by $w<-1$ and therefore by the violation of the weak energy condition. In several works it has been explored the theoretical possibilities of an expanding universe dominated by phantom dark energy. The phantom dark energy models \cite{caldwell} lead to a different types of singularities \cite{sergei21, bamba, dsaez}, the most drastic of which is the so called Big Rip singularity \cite{frampton}, \cite{kamion}, in which the scale factor, the density and pressure become infinite at a finite time. The type II or sudden singularities are characterized by finite $a$ and $\rho$ but divergent $p$ at finite time \cite{barrow}, \cite{barrow1}. In the type III singularities both $\rho$ and $p$ diverge, but the scale factor remains finite at finite time \cite{sergei22}, \cite{stefancic}. Finite-time singularities
which are softer than the previous, are classified as type IV singularities in which the scale parameter remains finite and the density and pressure become zero or finite at finite time, but higher derivatives of the Hubble parameter are divergent \cite{sergei21}. Recently, new models known as Little Rip have been proposed, in which the DE increases with time as in the Big Rip solutions, but without future singularity \cite{frampton1}, \cite{brevik},\cite{granda}. Cosmological singularities in extended theories of gravity have been studied in \cite{saez}, \cite{dsaez1}, \cite{dsaez2}, \cite{dsaez3}, \cite{dsaez4}.\\
Most of the studied singularities so far, are obtained by introducing explicit time dependence of the scale factor, or considering phenomenological models with generalized EoS in which the pressure density $p$ of dark energy is given in terms of the energy density as some function $p(\rho)$ \cite{sergei21}, \cite{barrow, barrow1, sergei22, stefancic, sergei25, barrow2, ruth, astashenok, jenkovszky}. The DE EoS for inhomogeneous fluid has been considered in \cite{bamba, nojiri}. A dark fluid unifying dark energy and dark matter has been proposed in \cite{chavanis1, chavanis2}, and a cosmological fluid with a turbulent component leading to Big Rip or Little Rip has been proposed in \cite{brevik2}. In \cite{capozziello1} a barotropic factor for the matter was proposed, which provides the negative pressure necessary for the accelerated expansion. In \cite{sdodintsov} a logarithmic-corrected power-law equation of state for the DE fluid was studied.  
In homogeneous and isotropic backgrounds like the FRW, another interesting way to construct cosmological models is to define the pressure density through the scale factor. This allows to incorporate the early time matter dominance in the general solution, giving rise among others, to phenomenologically viable cosmologies with phantom dark energy. In the present work we investigate new DE cosmologies with finite-time singularities, by introducing the pressure density as a function of the scale factor. These models present singularities of the type I known as Big-Rip, and of the type II-IV characterized by finite scale factor at finite time. The violation of strong or weak energy conditions becomes clear from the behavior of the EoS parameter. Explicit numerical examples showing acceptable behavior of the equation of state are given. In section 2 we present the background equations and give an example of DE with constant EoS, that leads to Big Rip singularity. In section 3 we consider models with type II-IV singularities with explicit numerical examples describing viable cosmological scenarios, and evaluate the time remaining to singularities. Some discussion is given in the final section.
\section{Basic equations}
We will consider a fluid in the spatially flat homogeneous and isotropic FRW background
\be\label{eq1}
ds^2=-dt^2+a(t)^2\sum_{i=1}^3dx_i^2
\ee
the dynamics is determined from the equations
\be\label{eq2}
H^2=\frac{\kappa^2}{3}\rho
\ee
\be
-3H^2-2\dot{H}=\kappa^2 p
\ee
where $\rho$ and $p$ are the energy and pressure densities of the fluid that makes up the universe. From these equations follows the continuity equation
\be\label{eq3}
\dot{\rho}+3H\left(\rho+p\right)=0
\ee
Another way to integrate this equation is by giving the pressure in terms of the scale factor, i.e. $p=p(a)$. As a result of solving the Eq. (\ref{eq3}) we find the density in terms of the scale factor, and therefore determine the Hubble parameter from (\ref{eq2}) in terms of the redshift. The scale factor can be found by integrating eq. (\ref{eq2}), which gives the time dependence of the scale factor and completes the cosmological description of the model. The Eq. (\ref{eq3}) may be written in terms of the scale factor as independent variable, as
\be\label{eq4}
a\frac{d\rho(a)}{da}+3\left[\rho(a)+p(a)\right]=0
\ee
Note that the general solution to this differential equation includes the solution to the homogeneous equation, corresponding to $p=0$. From Eqs. (\ref{eq3}) or (\ref{eq4}) follows that the limit $\dot{\rho}\rightarrow 0$ is reached as $p\rightarrow -\rho$, equivalently to $w\rightarrow -1$, which imply that $\ddot{\rho}\rightarrow 0$ and the EoS never crosses the phantom divide. Nevertheless if one adds a second fluid with constant EoS, $w=0$ (which corresponds to the solution to homogeneous equation, i.e. with $p=0$),
then the crossing of the phantom divide is possible for some models with finite scale factor singularities. 
So if we assume that $p(a)$ describes only the dark energy pressure, then the particular solution to Eq. (\ref{eq4}) gives the dark energy density as a function of the scale factor, giving in this approach the pressure and energy densities parametrically through the scale factor as the parameter. The scalar field responsible for the dark energy component can be found from the expressions for the density and pressure for minimally coupled quintessence (phantom) scalar field as follows 
\be\label{eq3a}
\rho=\pm \frac{1}{2}\dot{\phi}^2+V(\phi)
\ee
\be\label{eq3b}
p=\pm \frac{1}{2}\dot{\phi}^2-V(\phi)
\ee
which give
\be\label{eq3c}
\rho+p=\pm\dot{\phi}^2
\ee
where the minus sign corresponds to the phantom scalar. In therms of the scale factor, this equation can be written as
\be\label{eq3d}
\frac{d\phi}{da}=\pm \frac{1}{aH}\sqrt{|\rho(a)+p(a)|}
\ee
for the scalar field dominated universe, we find (using (\ref{eq2})
\be\label{eq3e}
\phi=\phi_0\pm \frac{\sqrt{3}}{\kappa}\int_{a_0}^a \frac{\sqrt{|\rho(a)+p(a)|}}{a\sqrt{\rho}} da.
\ee
From the Eqs. (\ref{eq3a}) and (\ref{eq3b}) we find the scalar potential as function of the scale factor
\be\label{eq3f}
V=\frac{1}{2}\left[\rho(a)-p(a)\right]
\ee
where the scalar field density and pressure satisfy the continuity equation (\ref{eq4}) which is equivalent to the equation of motion for the scalar field. So, the scalar field can be reconstructed from the given dark energy density and pressure in terms of the scale factor. In some cases it is possible to find the explicit expression for the potential in terms of the scalar field, but in any case it is always possible the numerical reconstruction of the scalar model. In the next sections we present several examples of models that give rise to different types of finite time future singularities.\\
\subsection*{The classical Big Rip singularity}
Let's consider the following simple model with $a$-dependence of the pressure density on the scale factor
\be\label{eq1a}
p=-\frac{A}{a^{\alpha}}
\ee
where $A$ is a positive constant. Replacing in Eq. (\ref{eq4}) we find
\be\label{eq1b}
\rho=\frac{3A}{(3-\alpha)a^{\alpha}}
\ee
where $\alpha<3$, and this solution gives the dark energy density. This solution corresponds to known power-law behavior of the scale factor which leads to constant equation of state given by
\be\label{eq1b1}
w_{DE}=-1+\frac{\alpha}{3}
\ee
The scale factor behaves as  
$$a(t)=a_0 t^{2/\alpha},$$
where $\alpha=0$ corresponds to the cosmological constant, $0<\alpha<2$ corresponds to quintessence dark energy ($-1<w_{DE}<-1/3$), and $\alpha<0$ describes phantom dark energy $w_{DE}<-1$. From Eqs. (\ref{eq3e}) and (\ref{eq3f}) we can find the corresponding scalar field and potential as (using $\kappa^2=M_p^{-2}$)
\be\label{eq1b2}
\phi=\phi_0\pm \sqrt{|\alpha|}M_p \ln\frac{a}{a_0}
\ee
\be\label{eq1b3}
V=\frac{(6-\alpha)A}{2(3-\alpha)}\frac{1}{a^{\alpha}}=\frac{(6-\alpha)A}{2(3-\alpha)a_0^{\alpha}}\exp\left[\pm \frac{\sqrt{|\alpha|}}{M_p}(\phi-\phi_0)\right]
\ee
which are the known scalar field and potential for the power-law expansion.
Taking the solution (\ref{eq1b}) for the dark energy density on can write the Hubble parameter, from the Friedmann equation, as 
\be\label{eq1b4}
H^2=\frac{\kappa^2}{3}\left[\frac{3A(1+z)^{\alpha}}{(3-\alpha)}\right]
\ee
If $\alpha<0$, then the EoS, according to (\ref{eq1b1}), will be bellow the phantom divide ($w_{DE}<-1$) leading in the future to blow up of the scale factor and curvature invariants in what is called future Big Rip (BR) singularity. In general, the time remaining to the BR singularity, if there is any, can be expressed as follows
\be\label{eq1b5}
ts-t_0=\int_1^{\infty}\frac{da}{aH}
\ee
where $t_0$ is the current time corresponding to $a=1$ or ($z=0$).
From the expression (\ref{eq1b4}) we can find the remaining time to the BR singularity as
\be\label{eq1b6}
t_{BR}=t_0+H_0^{-1}\int_{-1}^0 \frac{dz}{(1+z)(1+z)^{\alpha/2}}=t_0+2|\alpha|^{-1}H_0^{-1}
\ee
where we used the normalization $\frac{\kappa^2 A}{(3-\alpha)H_0^2}=1$. Note that the BR singularity is reached when $a\rightarrow\infty$ or $z$ takes the value $z=-1$ at finite time. 
One the other hand, one can also include the solution to the homogeneous equation when solving the continuity equation (\ref{eq4}), giving the following expression for the Hubble parameter 
\be\label{eq1c}
H^2=\frac{\kappa^2}{3}\left(\rho_{m0}(1+z)^3+\frac{3A(1+z)^{\alpha}}{(3-\alpha)}\right)
\ee
where $\rho_{m0}$ is an integration constant, giving the term which is associated with the pressure-less matter component (baryonic and dark matter). The above expression can be normalized using the flatness condition, as
\be\label{eq1c1}
H^2=H_0^2\left[\Omega_{m0}(1+z)^3+(1-\Omega_{m0})(1+z)^{\alpha}\right]
\ee
where,
\be\nonumber
\Omega_{m0}=\frac{\kappa^2 \rho_{m0}}{3H_0^2}, \;\;\; 1-\Omega_{m0}=\frac{\kappa^2 A}{(3-\alpha)H_0^2}
\ee
By flatness condition we mean that the current value ($z=0$) of the total density parameter is equal to 1, which applies to the flat FRW background given in Eq. (1). To find the effective equation of state, which includes the contribution from the dust-like matter component in addition to the dark energy component, we use (\ref{eq1a}) and (\ref{eq1c1}), which give the effective EoS in terms of the redshift as
\be\label{eq1d}
w_{eff}=\frac{p}{\rho}=-1-\frac{2\dot{H}}{3H^2}=-\frac{(3-\alpha)A}{(3-\alpha)\rho_{m0}(1+z)^{3-\alpha}+3A}
\ee
This expression has the correct asymptotic limits as expected, i.e. $w\rightarrow 0$ as $z\rightarrow\infty$, $w\rightarrow-(3-\alpha)/3$ as $z\rightarrow-1$, showing an appropriate behavior according to observations.  This means that if we consider that the actual value of $\alpha$ is negative, then necessarily in the future the effective EoS will cross the phantom divide and evolve towards the value $w(z\rightarrow-1)=-1+\alpha/3$. From (\ref{eq1c}) follows that (taking $\alpha<0$), in the future the universe becomes dominated by DE and the term proportional to $a^{-3}$ becomes negligible, giving rise to the discussed above Big Rip singularity. From (\ref{eq1c1}) and neglecting the matter term, we can find an approximate expression for the time from now to the BR singularity as $t_{BR}=t_0+2|\alpha|^{-1}H_0^{-1}(1-\Omega_{m0})^{-1/2}$, where $t_0$ is current time corresponding to $z=0$. The BR singularity here is obtained when $z$ reaches the value $z=-1$ at time $t_{BR}$. Note that the reconstruction of the scalar field using the pressure and energy densities (\ref{eq1a}) and (\ref{eq1b}) respectively, is valid for the case of dark energy dominated universe. \\
\noindent As a second case, let's consider the following pressure density
\be\label{eq1h}
p=-Aa^{\beta}e^{-1/a^{\gamma}}
\ee
At the limit $\gamma\rightarrow 0$, this pressure gives the power-law behavior of the previous case. Then, if $\gamma<<1$, this pressure could describe a departure form the power-law behavior (provided $\beta>-3$). Note also that if $\gamma>>1$, then in the future ($a>1$) the factor  $e^{-1/a^{\gamma}}$ can be neglected and the behavior becomes close  to the power-law.  
Integrating the Eq. (\ref{eq4}) we find the particular solution
\be\label{eq1i}
\rho=\frac{3A}{\gamma a^3}\Gamma[-\frac{\beta+3}{\gamma},a^{-\gamma}]
\ee
 $\Gamma[s,x]$ is the incomplete gamma function. Here we assume that $0<\beta<1$ and $0<\gamma<1$, which guarantees the following behavior: if $a\rightarrow\infty$, then $p\rightarrow-\infty$ and $\rho\rightarrow\infty$, giving rise to Big Rip singularity if this happens in finite time. 
The EoS for this dark energy fluid is given, in terms of the redshift, by
\be\label{eq1w}
w_{DE}=-\frac{a^{\beta+3}e^{-1/a^{\gamma}}}{3\Gamma[-\frac{\beta+3}{\gamma},a^{-\gamma}]}
\ee
In Fig. 1 we show the evolution of $w_{DE}$ as function of the redshift
\begin{center}
\includegraphics [scale=0.6]{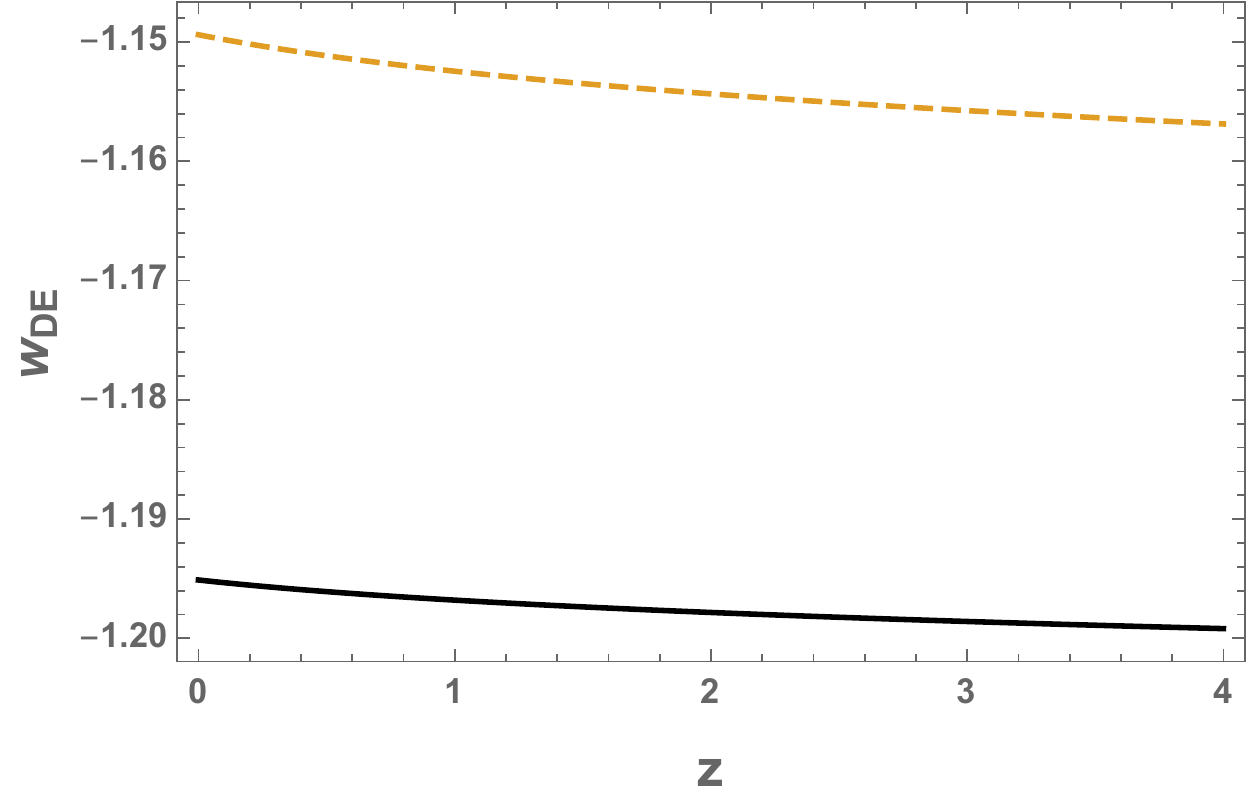}
\end{center}
\begin{center}
{Fig. 1 \it The DE equation of state parameter $w_{DE}$ versus redshift, for $\beta=1/2, \gamma=1/12$ (continuous line) and  $\beta=1/3, \gamma=1/9$ (dashed line). In the far future, at $z\rightarrow -1$, the curves evolve toward $-7/6$  (continuous line) and $-10/9$ (dashed line).} 
\end{center}
Assuming a DE dominated universe, the time remaining to the singularity can be evaluated as
\be\label{eq1j}
\begin{aligned}
t_s-t_0=&H_0^{-1}\int_1^{\infty}a^{-1}\left(\frac{A}{\gamma H_0^2 M_p^2}\right)^{-1/2}\times \\ &
\left[ \frac{1}{a^3}\Gamma[-\frac{\beta+3}{\gamma},a^{-\gamma}]\right]^{-1/2}da
\end{aligned}
\ee
where $t_0$ is the current time corresponding to the scale factor $a=1$. Note that in this integral we may express the constant $A$ in units of $H_0^2M_p^2$, where $H_0$ is the current Hubble parameter. To show that $t_s$ takes finite value, let's consider the numerical values of Fig. 1, Assuming $A=2.5$. For $\beta=1/2, \gamma=1/12$, it is found that at $a\rightarrow\infty$, $w_{DE}\rightarrow -7/6$ and the time remaining to the BR singularity is $t_s-t_0\approx 7H_0^{-1}$. In the second case, for $\beta=1/3, \gamma=1/9$,  the DE EoS $w_{DE}\rightarrow -10/9$ at $a\rightarrow \infty$ and the time remaining to the BR singularity is $t_s-t_0\approx 9.5H_0^{-1}$. \\
The corresponding scalar field can be found from (\ref{eq3e}) as 
\be\label{eq1i1}
\phi=\phi_0\pm \sqrt{3}M_p\int_{a_0}^a \frac{\sqrt{|\frac{3A}{\gamma a^3}\Gamma[-\frac{\beta+3}{\gamma},a^{-\gamma}]-Aa^{\beta}e^{-1/a^{\gamma}}|}}{a\sqrt{\frac{3A}{\gamma a^3}\Gamma[-\frac{\beta+3}{\gamma},a^{-\gamma}]}} da
\ee
and using (\ref{eq3f}), the potential is given by the following expression
\be\label{eq1i2}
V=\frac{1}{2}\left[\frac{3A}{\gamma a^3}\Gamma[-\frac{\beta+3}{\gamma},a^{-\gamma}]+Aa^{\beta}e^{-1/a^{\gamma}}\right]
\ee
The integration in (\ref{eq1i1}) can not be performed analytically, but these equations allow to follow numerically the behavior of the scalar potential in terms of the scalar field, for a given cosmological epoch, represented by an interval of the scale factor or a redshift interval.
 
\noindent By including the homogeneous solution to the continuity equation (\ref{eq4}) we find the following expression for the Hubble parameter, that includes the contribution form the dust-like matter component
\be\label{eq1k}
H^2=\frac{\kappa^2}{3}\left(\rho_{m0}(1+z)^3+\frac{3A}{\gamma a^3}\Gamma[-\frac{\beta+3}{\gamma},a^{-\gamma}]\right)
\ee
Taking the appropriate value $\rho_{m0}=0.3$ and applying the flatness condition at $a=1$, to the Friedmann equation (\ref{eq1k}) we find $A\simeq 2.27$, for $\beta=1/2, \gamma=1/12$, and $A\simeq 2.46$ for $\beta=1/3, \gamma=1/9$. Using these values and the expression (\ref{eq1b5}), we find the time remaining to the Big Rip singularity as: $t_s-t_0\approx 7.39 H_0^{-1}$ for the first case and $t_s-t_0\approx 9.67 H_0^{-1}$ for $A\simeq 2.46$. The behavior of the effective EoS (including the matter component) is shown in Fig. 2.
\begin{center}
\includegraphics [scale=0.6]{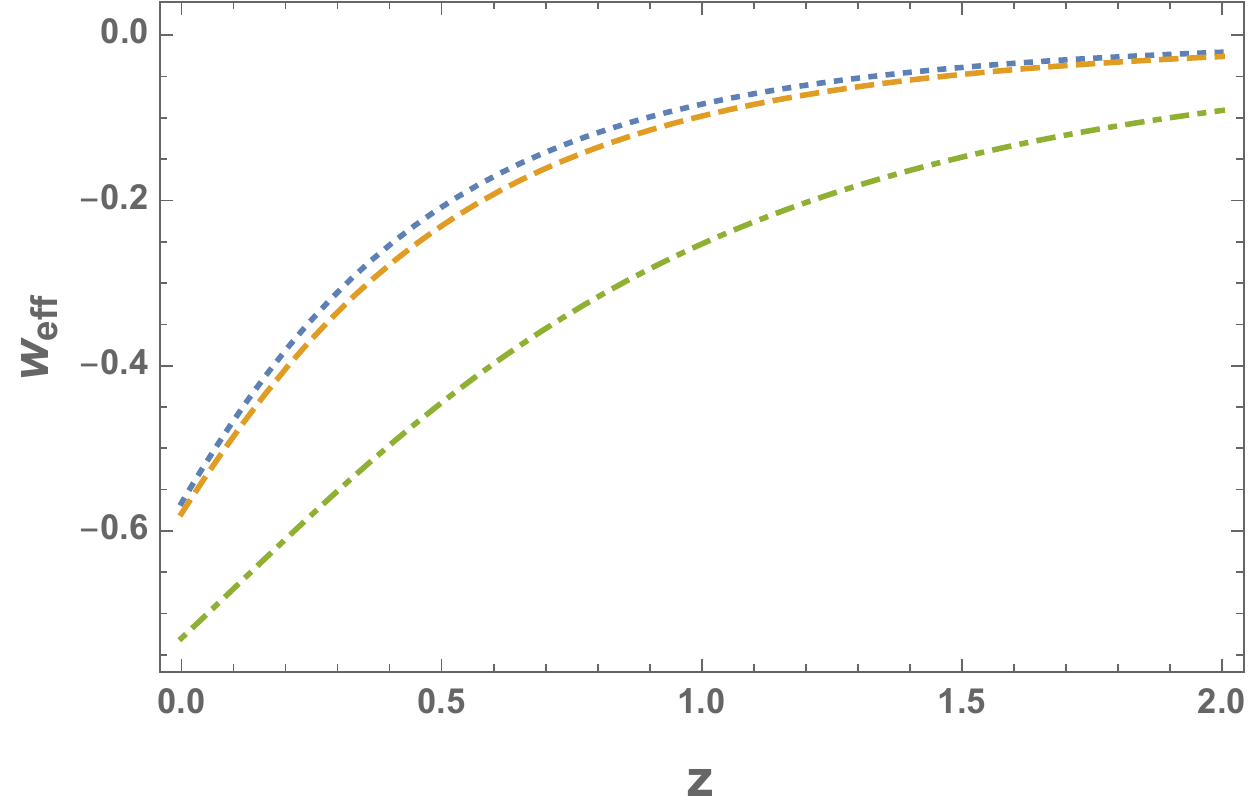}
\end{center}
\begin{center}
{Fig. 2 \it The effective equation of state parameter $w$ versus redshift, for the parameters $\beta=1/2, \gamma=1/12, A=2.27$ and  $\beta=1/3, \gamma=1/8, A=2.46$. These values are subject to the flatness condition, with $\Omega_{m0}=0.3$. The bottom curve corresponds to the $\Lambda$CDM model.} 
\end{center}
\section{Future singularities with finite scale factor}
In this section we consider dark energy models with future singularities at finite time and finite scale factor. We will propose different models with pressure $p(a)$ that lead to singularities classified as of type II-IV. The resulting cosmologies give acceptable description of the evolution from the early matter dominance to the current time characterized
by predominance of the dark energy, whose EoS is quite consistent with being below the phantom divide.
\subsection*{Type III singularities}
Let's consider the following $a$-dependence of the density on the scale factor
\be\label{eq5}
p=-\frac{A}{a^{\alpha}}-\frac{B}{(a_s-a)^{\beta}}
\ee
where $A$ and $B$ are positive constants and $a<a_s$. Replacing in Eq. (\ref{eq4}) we find
\be\label{eq6}
\begin{aligned}
&\rho=\frac{3A}{(3-\alpha)a^{\alpha}}+\\ &\frac{3B[2a_s^2-2a_s (\beta-1)a+(\beta-1)(\beta-2)a^2](a_s-a)^{1-\beta}}{(\beta-1)(\beta-2).(\beta-3)a^3}
\end{aligned}
\ee
Making $B=0$ gives the already considered model with constant equation of state and power-law expansion. The second term represents small correction (assuming $B<<A$) to the power-law scenario for the evolution from early times to the present, or in any case for times away from the singularity at $a_s$. Close to the singularity at $a=a_s$, in the future, the second term dominates giving rise to type III singularity as shown below. From (\ref{eq5}) and (\ref{eq6}) follows the dark energy  equation of state (EoS)
\be\label{eq7}
w_{DE}=-\frac{a^3\left[A(a_s-a)^{\beta}+Ba^{\alpha}\right]}{(a_s-a)^{\beta}\left(\frac{3A}{3-\alpha}\right)a^3+3Bf(a)a^{\alpha}(a_s-a)}
\ee
where 
\be\label{eq8}
f(a)=\frac{(\beta-1)(\beta-2)a^2-2(\beta-1)a_s a+2a_s^2}{(\beta-1)(\beta-2)(\beta-3)}
\ee
Note that at the limit $a\rightarrow 0$ the energy density becomes dominated by the terms $\rho\propto a^{-3}$, which is typical of early time cosmology, characterized by tracking behavior of the dark energy and dominance of matter. To analyze the cosmological implications of this solution, first note that to keep $\rho\geq 0$ we have to consider the appropriate interval for $\beta$. Thus we exclude the values $\beta<1$ and $2<\beta<3$, as in the first case the last term in (\ref{eq6}) becomes negative and in the second case the last term in (\ref{eq6}) changes the sign before $a$ reaches the value $a_s$. So we will consider the intervals $1<\beta<2$ and $\beta>3$.\\
In that case, we see that as $a \rightarrow a_s$, $p\rightarrow-\infty$, $\rho\rightarrow\infty$, and from (\ref{eq7}) follows $w\rightarrow-\infty$. This behavior corresponds to type III singularity \cite{sergei22}, \cite{stefancic}. Despite this singularity the model behaves in good agreement with observations. In fact, from (\ref{eq7}) (assuming that $a=0$ corresponds to $t=0$) we see that as $a\rightarrow 0$ the EoS $w\rightarrow 0$. Using (\ref{eq6}) and introducing the current value of the Hubble parameter $H_0$, the Friedmann equation may be written as
\be\label{eq9}
\begin{aligned}
H^2=& H_0^2\Big[\Omega_{\alpha}a^{-\alpha}+\left(\Omega_{\beta 1}a^{-3}+\Omega_{\beta 2}a^{-2}+\Omega_{\beta 3}a^{-1}\right)\times \\& \left(\frac{a_s-a}{a_s-1}\right)^{1-\beta}\Big]
\end{aligned}
\ee
where
\be\label{eq10}
\begin{aligned}
&\Omega_{\alpha}=\frac{\kappa^2 A}{(3-\alpha)H_0^2},\,\,\,\, \Omega_{\beta 1}=\frac{2\kappa^2 B a_s^2(a_s-1)^{1-\beta}}{(\beta-1)(\beta-2)(\beta-3)H_0^2}\\
&\Omega_{\beta 2}=-\frac{2\kappa^2 B a_s(a_s-1)^{1-\beta}}{(\beta-2)(\beta-3)H_0^2},\,\,\,\, \Omega_{\beta 3}=\frac{\kappa^2 B(a_s-1)^{1-\beta}}{(\beta-3)H_0^2}
\end{aligned}
\ee
where we have normalized the expression in such a way that $H(a=1)=H_0$, with $a=1$ being the current value of the scale factor. This leads to the flatness condition at $a=1$
\be\label{eq11}
\Omega_{\alpha}+\sum_{i=1}^3\Omega_{\beta i}=1
\ee
Note that the coefficients $\Omega_{\beta1,2,3}$ in (\ref{eq9}) can  not be interpreted as kind of dust, dark curvature and topological defects density parameters respectively, due to the $a$-dependent  common factor $(a_s-a)^{1-\beta}$. Nevertheless, if $a_s>>1$ and away from the singularity the factor $(\frac{a_s-a}{a_s-1})^{1-\beta}\sim 1$ and the terms with $\Omega_{\beta1,2,3}$ behave almost like the aforementioned types of density. \\
\noindent In Fig. 3 we show the behavior of the DE EoS for three sets of constants, subject to the flatness restriction (\ref{eq11}). The values of  $A$ and $B$ are given in units of $3H_0^2/\kappa^2=3H_0^2M_p^2$. 
\begin{center}
\includegraphics [scale=0.6]{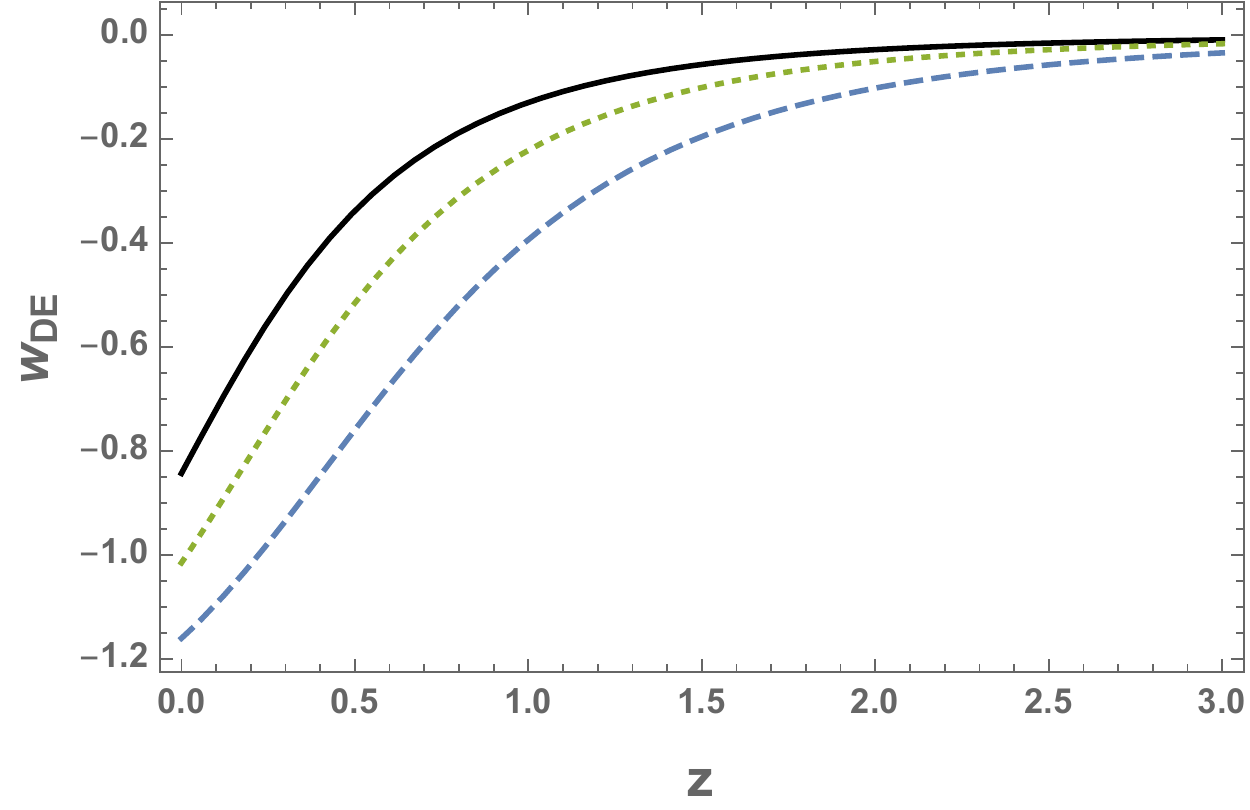}
\end{center}
\begin{center}
{Fig. 3 \it These curves describe the DE EoS parameter $w$ versus redshift, for $\alpha=-1, \beta=3/2, B=2.30\times 10^{-5}$ and the three future scenarios $(a_s=50, A=1.16)$, $(a_s=75, A=1.01)$, $(a_s=100, A=0.84)$ (subject to the flatness condition), corresponding to the dashed, dotted and continuous curves respectively. The dotted curve shows appropriate scaling behavior at early times and is currently close to the cosmological constant. The upper curve behaves closely as an effective EoS unifying DE and dark matter.}
\end{center}
It can be shown that the curves in Fig. 3 are more sensitive to the changes in $A$, i.e.  given $A$ the EoS for $a_s=100,50,20,10$ are practically indistinguishable from each other before the singularities. These curves show behavior similar to the dust matter dominance at early times, and evolve trough three different phases of decelerated, accelerated and phantom expansion. So the model with the density and pressure given parametrically by (\ref{eq5}) and (\ref{eq6}) can give a unified description of the dark matter and dark energy, including also the, not ruled out, phase of phantom expansion. Note that even though the pressure (\ref{eq5}) and density (\ref{eq6}) satisfy the continuity equation, the DE EoS crosses the phantom divide. This is because each term in the pressure and the two terms in the density (the one that depends on $A$ and $\alpha$ and the one that depends on $B$ and $\beta$) behave as two independent fluids, each of one satisfying the continuity equation. The scalar field representation in this case requires two scalar fields, one of which should be phantom.\\
\noindent One can also include the matter content in the Friedmann equation, as follows if one includes the homogeneous solution to Eq. (\ref{eq4}), which adds the term $\rho_{m0}/a^3$ to the density (\ref{eq6}), being $\rho_{m0}$ the constant of integration. Then, the Friedmann equation becomes 
\be\label{eq9a}
\begin{aligned}
H^2=&H_0^2\Big[\Omega_{mo}a^{-3}+\Omega_{\alpha}a^{-\alpha}+\\ &\left(\Omega_{\beta 1}a^{-3}+\Omega_{\beta 2}a^{-2}+\Omega_{\beta 3}a^{-1}\right)\left(\frac{a_s-a}{a_s-1}\right)^{1-\beta}\Big]
\end{aligned}
\ee
where $$\Omega_{m0}=\frac{\kappa^2\rho_{m0}}{3H_0^2}$$
and the other density parameters have been defined in (\ref{eq10}). The flatness condition becomes
\be\label{eq11a}
\Omega_{m0}+\Omega_{\alpha}+\sum_{i=1}^3\Omega_{\beta i}=1,
\ee
and the effective EoS takes the form
\be\label{eq7a}
w_{eff}=\frac{-a^3\left[A(a_s-a)^{\beta}+Ba^{\alpha}\right]}{(a_s-a)^{\beta}\left[\rho_{m0}a^{\alpha}+\frac{3A}{3-\alpha}a^3\right]+3Bf(a)a^{\alpha}(a_s-a)}
\ee
In Fig. 4 we plot the evolution of the effective EoS for the same scenario considered in Fig. 3, but taking into account the flatness condition given by (\ref{eq11a}). The parameters $\rho_{m0}, A, B$ are given in units of $3H_0^2M_p^2$.
\begin{center}
\includegraphics [scale=0.6]{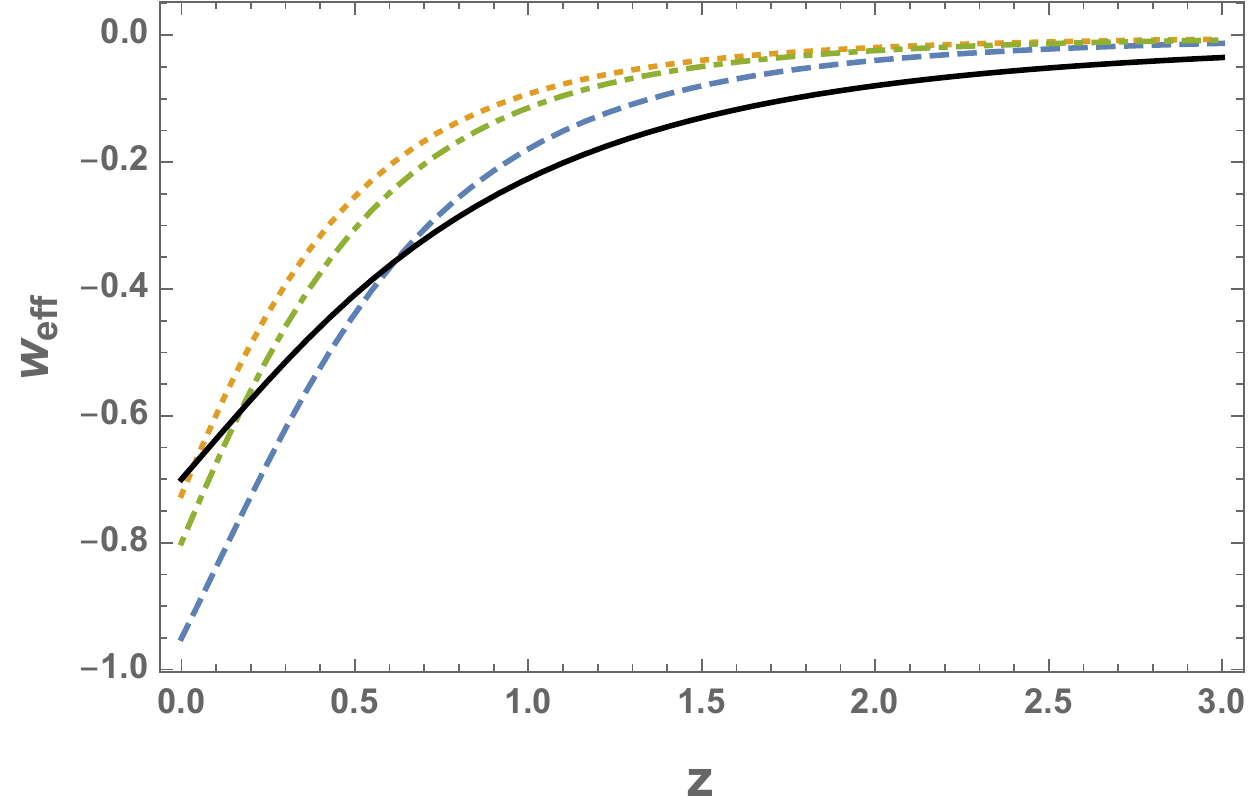}
\end{center}
\begin{center}
{Fig. 4 \it The effective EoS parameter versus the redshift, for $\alpha=-1, \beta=3/2, B=2.30\times 10^{-5}$ and the three future scenarios $(a_s=50, A=1.16)$, $(a_s=75, A=1.01)$, $(a_s=100, A=0.84)$ (subject to the flatness condition (\ref{eq11a})), corresponding to the dashed, dotted and continuous curves respectively. The dotted curve shows appropriate scaling behavior at early times and is currently close to the cosmological constant. The solid line corresponds to the $\Lambda$CDM model. }
\end{center}
Using Eq. (\ref{eq9}) we can find the time to the type III singularity, starting from the current time $t_0$ corresponding to $a=1$. From (\ref{eq9}) follows
\be\label{eq12}
\begin{aligned}
&t_s-t_0=H_0^{-1}\int_1^{a_s}a^{-1}\Big[\Omega_{mo}a^{-3}+\Omega_{\alpha}a^{-\alpha}+\\ &\left(\Omega_{\beta 1}a^{-3}+\Omega_{\beta 2}a^{-2}+\Omega_{\beta 3}a^{-1}\right)\left(\frac{a_s-a}{a_s-1}\right)^{1-\beta}\Big]^{-1/2}da
\end{aligned}
\ee
In fact this time is finite, as the sub-integral function remains bounded along the interval of integration, since for $1<\beta<2$ or $\beta>3$ 
and at the limit $a\rightarrow a_s$, the sub-integral function behaves as $(a_s-a)^{(\beta-1)/2}\rightarrow 0$. For the three examples shown in Fig. 4, the times remaining before the type III singularity are about $2.1 H_0^{-1}$, $2.4 H_0^{-1}$ and $2.8 H_0^{-1}$, for $a_s=50, 75, 100$ respectively.\\
It is worth mentioning that if there is any way of overcoming the type III singularity and the universe could continue the expansion, then in the present model there is one possibility: if $\beta$ is an odd integer and $\beta>3$, then the energy density remains positive even beyond the singularity at $a>a_s$ (other values of $\beta$ are not allowed physically in this case as the density would be negative or imaginary) and the universe continue evolving in the phantom phase towards the state with $w=-(1-\alpha/3)$. However as can be seen from (\ref{eq9}), as the scale factor continue increasing beyond $a_s$, all negative powers of $a$ fall sufficiently fast and become irrelevant leading to the dominance of the remaining DE term $\Omega_{\alpha} a^{-\alpha}$. \\

\subsection*{Sudden singularities}
The sudden singularity is a finite time singularity that occurs when the scale factor $a(t)$, its time derivative and the energy density $\rho$ remain finite, while the pressure $p\rightarrow \infty$, and therefore, the EoS also suffers the singularity, $w\rightarrow-\infty$. To illustrate this type of singularities we consider the following pressure density
\be\label{eq1f}
p=-\frac{A a^{\beta}}{(a_s-a)^{\alpha}}
\ee
where $A$ is a positive constant and $a<a_s$. Integrating the continuity equation (\ref{eq4}), with the pressure density given by (\ref{eq1f}), we find a general solution which depends on the hyper geometric function
\be\label{eq12a}
\rho=\frac{3Aa^{\beta}}{(3+\beta)a_s^{\alpha}}\hspace{0.1cm} _{2}F_1\left[3+\beta,\alpha,4+\beta,\frac{a}{a_s}\right].
\ee
But we will limit here to one of the cases that gives the result in terms of elementary functions, namely $\alpha=1/2$, $\beta=1/2$ which gives $p=-\frac{A a^{1/2}}{(a_s-a)^{1/2}}$. Integrating (\ref{eq4}) for this pressure gives the energy density
\be\label{eq14}
\begin{aligned}
\rho(a)=&\frac{A}{8}\Big[15\left(\frac{a_s}{a}\right)^3\arcsin\left(\sqrt{\frac{a}{a_s}}\right)-\\ &\sqrt{\frac{a_s}{a}-1}\left(8+10\frac{a_s}{a}+15\left(\frac{a_s}{a}\right)^2\right)\Big]
\end{aligned}
\ee
The dark energy EoS is
\be\label{eq15}
w_{DE}=-\frac{A\left(\frac{a_s}{a}-1\right)^{-1/2}}{\rho(a)}
\ee
with $\rho(a)$ given by (\ref{eq14}).
At the limit $a\rightarrow a_s$, from (\ref{eq1f}) (for $\alpha=1/2$, $\beta=1/2$) and (\ref{eq14}) follows that 
\be\label{eq16}
\rho\rightarrow\frac{15}{16}A\pi,\,\,\, p\rightarrow -\infty,\,\,\,\, w\rightarrow-\infty
\ee
giving rise to sudden singularity.\\
Taking into account that the pressure for $\alpha=\beta=1/2$ takes the form
\be\label{eq1gw}
p=-\frac{A\sqrt{a}}{\sqrt{a_s-a}},
\ee
one can eliminate $a_s/a$ in (\ref{eq14}) and (\ref{eq1gw}) and obtain the explicit relation between the density and pressure as
\be\label{eq1hx}
\begin{aligned}
\rho =& \frac{A}{8}\Big[\frac{A}{p}\left(15\left(1+\frac{A^2}{p^2}\right)^2+10\left(1+\frac{A^2}{p^2}\right)+8\right)-\\ &15\left(1+\frac{A^2}{p^2}\right)^3\arcsin\frac{p}{\sqrt{p^2+A^2}}\Big]
\end{aligned}
\ee
From this expression follow the same previous limits, 
\be\label{eqjy}
\lim_{p\to -\infty}\rho=\frac{15}{16}A\pi,\,\,\, \lim_{p\to -\infty}\frac{p}{\rho}=-\infty
\ee
It is interesting to note that in the present case the dark energy is always in the phantom phase.  This follows by taking the limits  $a\rightarrow0$ and $a\rightarrow a_s$ in (\ref{eq15}) giving $w_{DE}\rightarrow -7/6$ and $w_{DE}\rightarrow-\infty$ respectively. In Fig. 5 we show the dark energy EoS vs the redshift for some cases.
\begin{center}
\includegraphics [scale=0.6]{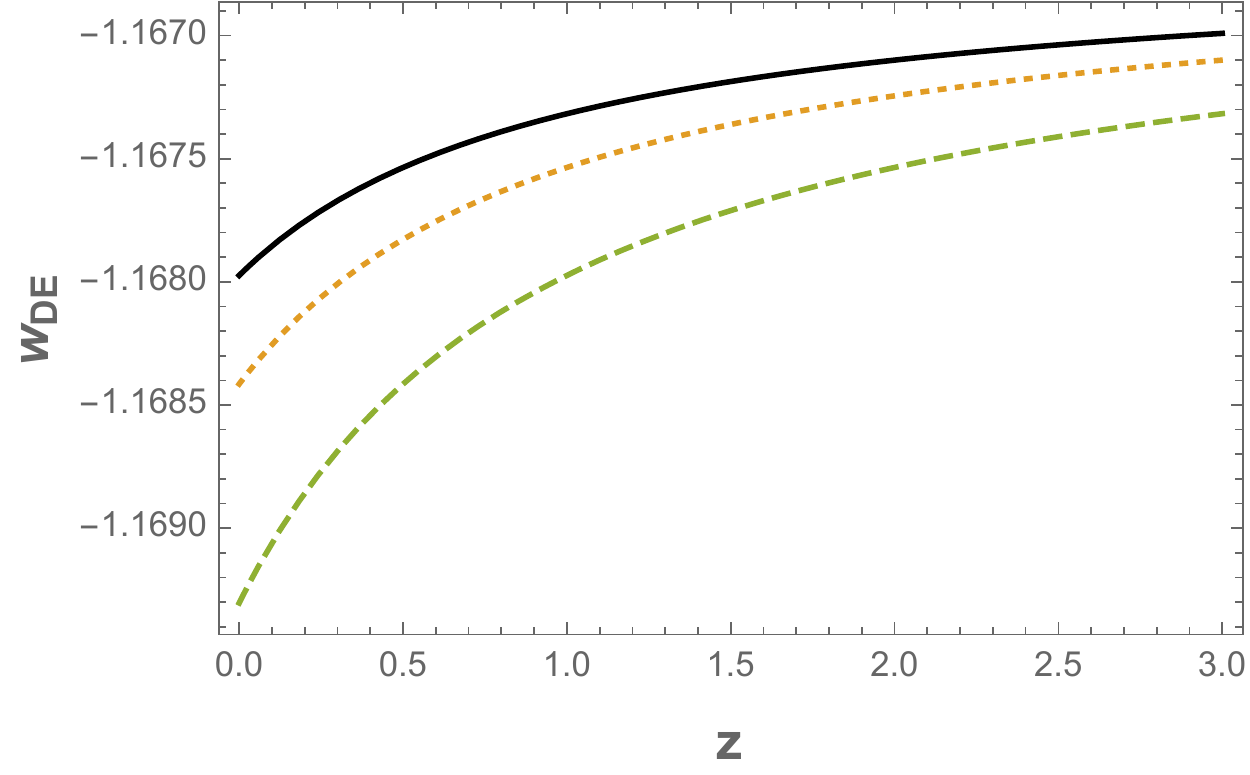}
\end{center}
\begin{center}
{Fig. 5 \it The DE EoS parameter $w_{DE}$ versus the redshift, for the evolutionary scenarios:  $a_s=100, A=11.62$ (solid line), $a_s=75, A=10.05$ (dotted line) and $a_s=50, A=8.2$ (dashed line) where the flatness condition, at $a=1$, has been taken into account.}
\end{center}
Adding the homogenous solution to (\ref{eq14}), one can write the total density as
\be\label{eq12b}
\rho=\frac{\rho_{m0}}{a^3}+\frac{3Aa^{\beta}}{(3+\beta)a_s^{\alpha}}\hspace{0.1cm} _{2}F_1\left[3+\beta,\alpha,4+\beta,\frac{a}{a_s}\right]
\ee
where $\rho_{m0}$ is identified with the matter component. At the limit $a\rightarrow a_s$ (assuming $\alpha=1/2, beta=1/2$) it follows 
\be\label{eq16a}
\rho\rightarrow\frac{\rho_{m0}}{a_s^3}+\frac{15}{16}A\pi,\,\,\, p\rightarrow -\infty,\,\,\,\, w\rightarrow-\infty
\ee
For $\alpha=\beta=1/2$, the effective EoS takes the form
\be\label{eq15a}
w_{eff}=-\frac{A\left(\frac{a_s}{a}-1\right)^{-1/2}}{\left[\frac{\rho_{m0}}{a^3}+\frac{A}{8}\left[15\left(\frac{a_s}{a}\right)^3\arcsin\left(\sqrt{\frac{a}{a_s}}\right)-\sqrt{\frac{a_s}{a}-1}\left(8+10\frac{a_s}{a}+15\left(\frac{a_s}{a}\right)^2\right)\right]\right]}
\ee
The evolution of $w_{eff}$ is shown in Fig. 6, for the cosmological scenarios ($a_s=50, 75, 100$) depicted in Fig. 5, taking into account the flatness condition for the density (\ref{eq12b}) ($A$ and $\rho_{m0}$ are given in units of $3M_p^2H_0^2$)
\begin{center}
\includegraphics [scale=0.6]{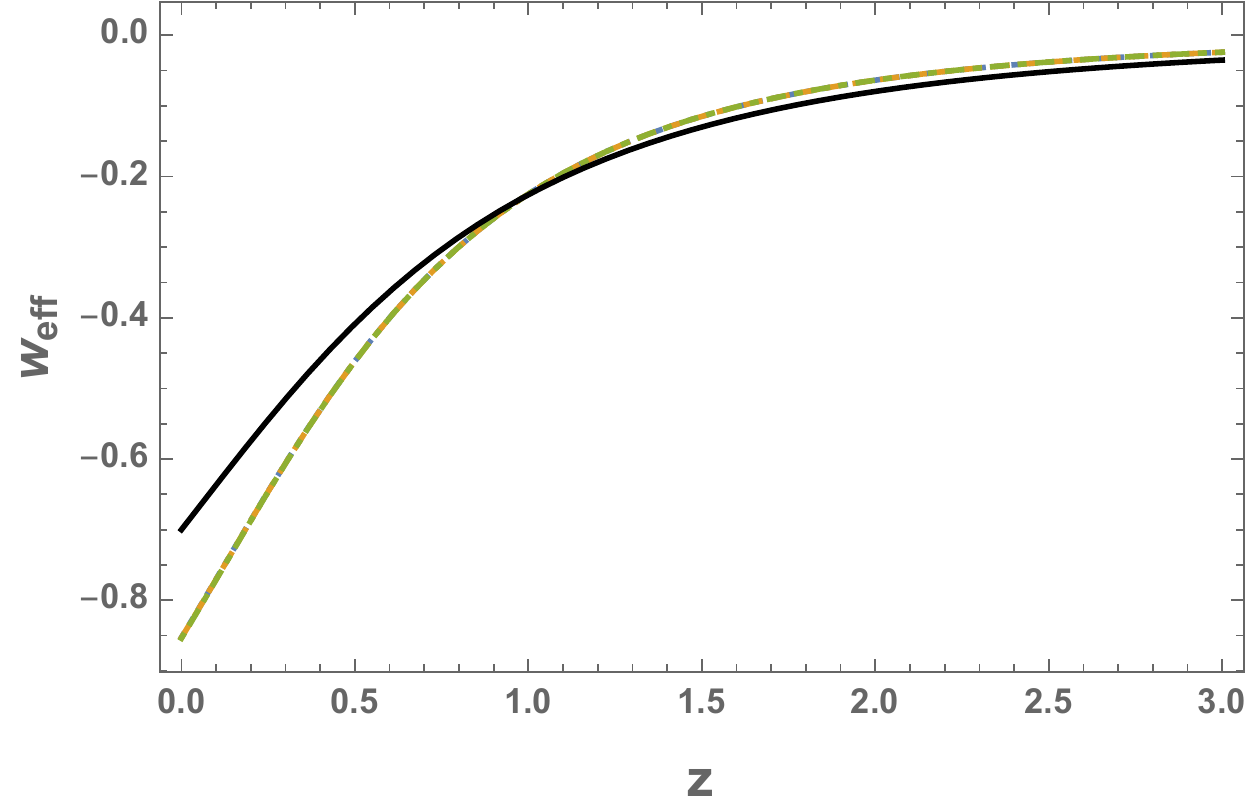}
\end{center}
\begin{center}
{Fig. 6 \it The effective EoS for the model (\ref{eq12b}) for the three cosmological scenarios: ($a_s=100, A=8.48$), ($a_s=75, A=7.34$)
and ($a_s=50, A=5.98$). Note that $w_{eff}$ is is very weakly sensitive to the time at which the singularity occurs (at least for the values $\alpha=1/2$ and $\beta=1/2$ considered in this case). It evolves toward phantom phase in the future and faces sudden singularity as $z\rightarrow z_s$. The solid line corresponds to the $\Lambda$CDM model.}
\end{center}
This EoS behaves as presureless matter at early times, presents the deceleration-acceleration transition in the expected redshift region, and evolves towards a phantom phase in the future, where it faces the sudden singularity.  
Using (\ref{eq14}) we find the time remaining to the singularity for the above three scenarios: $t_s-t_0\approx 2.7 H_0^{-1}$ for $a_s=50$, $t_s-t_0\approx 2.9 H_0^{-1}$ for $a_s=75$ and $t_s-t_0\approx 3 H_0^{-1}$ for $a_s=100$.\\
Considering other possible DE scenarios, one can see that, by integrating the equation (\ref{eq4}) for the pressure $p=-\frac{A a^{n}}{(a_s-a)^{1/2}}$, where $n$ is positive or negative integer, it can be directly checked that the DE density becomes negative as $A>0$, or if we try to satisfy the current flatness condition then $A$ should be negative which leads to $p>0$ and decelerated solution. This excludes integer values of $\beta$ in (\ref{eq12a}) for DE solutions corresponding to $\alpha=1/2$. In general, by integrating the equation (\ref{eq4}) for the pressure of the form $p=-\frac{A a^{(2n+1)/2}}{(a_s-a)^{1/2}}$, $(n=0,1,2,...)$, then the solutions present sudden singularities, but the current DE EoS becomes situated below the phantom barrier, and for larger $n$ the model becomes darker in the sense that the EoS moves away from the phantom barrier to more negative values. On the contrary if ($n=-1,-2,...$), then the DE EoS becomes situated above the phantom barrier despite the fact that the solutions present future sudden singularity, and the behavior of the EoS does not 
adjust well with the observations (the weak energy condition is preserved, and the strong energy condition could also be preserved before the singularity). On the other hand, if we consider pressures of the from $p=-\frac{A a^{1/n}}{(a_s-a)^{1/2}}$, $(n=2,3,...)$, then the resulting model behaves in good agreement with observations, and as $n$ increases, the behavior of the DE EoS becomes closer to that of the cosmological constant, but remaining bellow the phantom barrier, what might adjust better to the current data. And for negative $n$, as $|n|$ increases, the DE EoS approaches the cosmological constant from above. 
\subsection*{Type IV singularities}
Let's assume that $p$ has the form
\be\label{eq17}
p=-Aa^{\beta}(a_s-a)^{\alpha},\,\,\, -1<\beta<1, \beta\neq 0,\,\,\,\, 0<\alpha<1
\ee
Integrating Eq. (\ref{eq4}) for this pressure, we find
\be\label{eq17a}
\rho=\frac{3A a_s^{\alpha}a^{\beta}}{\beta+3}\hspace{0.1cm} _{2}F_1\left[\beta+3,-\alpha,\beta+4,\frac{a}{a_s}\right]
\ee

which leads to the following limits as $a\rightarrow a_s$:
\be\label{eq18}
\begin{aligned}
\rho\rightarrow &\frac{3A a_s^{\alpha+\beta}\Gamma[\alpha+1]\Gamma[\beta+4]}{(\beta+3)\Gamma[\alpha+\beta+4]},\,\,\, p\rightarrow 0,\,\, \\ &
|H^{(n)}|\rightarrow \infty, n=2,3,.....
\end{aligned}
\ee
where $H^{(n)}$ denote the $n$-th time-derivative of $H$. The last limit follows from the fact that the derivative of $p$ with respect to $a$ diverges at $a\rightarrow a_s$ for $0<\alpha<1$, and the same follows for higher derivatives. Thus, the model (\ref{eq17}) gives rise to type IV singularities where the density becomes finite and non-zero at $a_s$,  which leads to non-singular EoS at $a_s$. Under appropriate choice of the parameters that respect the flatness condition and assuming a value for $a_s$, all these models lead to appropriate cosmological evolution and give a current value of the DE EoS in the range obtained from observations. It follows from (\ref{eq17}) and (\ref{eq17a}) that for a given $\alpha$ the behavior of the model improves as $|\beta|$ becomes smaller. Thus for $-1<\beta<0$, as $\beta\rightarrow -0$ the DE EoS becomes closer to the cosmological constant, approaching the phantom barrier ($w=-1$) from above, and for $0<\beta<1$, as $\beta\rightarrow +0$ the DE EoS approaches the phantom barrier from bellow (in both cases the strong energy condition becomes broken). Note that if we change $\alpha$ by $\alpha+n$ ($n=1,2,...$) in (\ref{eq17}), then starting from $(n+2)$-th order all time-derivatives of the Hubble parameter diverge.
\section{Discussion}
In most theoretical models, the dark energy is due to the negative pressure of an unknown kind of matter, and one sees appropriate to introduce models of DE by defining parametrically, in terms of the scale factor, the pressure and density of the DE fluid.
We have proposed various fluid models for dark energy by introducing the pressure density $p(a)$ as a given function of the scale factor $a$, with the density $\rho(a)$ determined from the continuity equation. This allows to introduce new phenomenological models of dark energy, and among them, we considered models that contain finite time future singularities.  The model (\ref{eq1a}) leads to the widely studied power-law expansion, while the models (\ref{eq1h}) and (\ref{eq1f}) contain a small deviation from the power-law as can be seen in the behavior of the EoS in Figs. 1 and 5.  The model (\ref{eq5}) represents a kind of effective density since it contains a dark-matter like density, and kind of dark curvature and topological defects densities that can be interpreted as such in the case $a_s>>1$.
Two models with Big Rip singularity have been considered, for a pressures of the form given by Eqs. (\ref{eq1a}) and (\ref{eq1h}). In Figs. 1 and 2 it is shown the redshift evolution of the DE  EoS for the fluid  (\ref{eq1h}) and the respective effective EoS (which includes the matter content) for some parameters, with the appropriate flatness conditions. The DE EoS (Fig. 1) remains in the phantom phase with adequate values for DE, and the effective EoS (Fig. 2) behaves properly, with matter and DE dominance phases, including the future phantom phase leading to the Big Rip singularity.
The DE fluid (\ref{eq5}) with $1<\beta<2$ or $\beta>3$ describes a cosmology with future type III singularity where $p\rightarrow -\infty$ and $\rho\rightarrow\infty$ for finite $a=a_s$ as follows from (\ref{eq5}) and (\ref{eq6}). In this model the DE equation of state presents very interesting features, since it simulates scaling behavior at early times and evolves from quintessence phase to phantom phase in such a way that the current value of the EoS parameter is close to $-1$, or even below this value as shown in the numerical examples depicted in figure 3. 
The sudden singularity described by the model (\ref{eq14}) or (\ref{eq1hx}), gives an EoS in the phantom region with acceptable values, as shown in the numerical examples depicted in Fig. 5, which also lead to acceptable evolution of the effective EoS as shown in Fig. 6. 
In general, the pressures of the form $p=-\frac{A a^{\pm 1/n}}{(a_s-a)^{1/2}}$, $(n=2,3,...)$ lead to sudden singularities and well behaved cosmologies, where for large positive $n$ the DE EoS approaches the phantom barrier from bellow and for large negative $n$ the DE EoS approaches the phantom barrier from above. Finally we considered a set of models characterized by the product of $\alpha$-power of $(a_s-a)$ times $\beta$-power of $a$, subject to the conditions (\ref{eq17}), which give an acceptable description of the DE, and generate future type IV singularities characterized by finite non-zero density that leads to non-singular EoS. Besides type IV singularities, this model also describes more softer behaviors where, starting from some n-th order, all derivatives of the Hubble parameter diverge.\\
Resuming, in the present work we investigated some possible models of dark energy fluid in the homogeneous and isotropic FRW background, where the pressure density was defined as a function of the scale factor. 
The numerical examples indicate that all the models considered above can support the current observations of accelerated expansion. We have shown explicit examples of dark energy that realize Big Rip, sudden, type III and type IV singularities \cite{sergei21}. Is worth noting that all models reproduce essentially the same expansion
law up to the present, despite the fact that they have different future evolution. Thus, in the case the universe ends in a future singularity, the current data make it difficult to determine what kind of singularity will take place in the future.
\section*{Acknowledgments}
This work was supported by Universidad del Valle under project CI 71074 and by COLCIENCIAS grant number 110671250405.

\nocite{*}

\end{document}